\documentclass[12pt]{article}
\usepackage{amssymb,amsmath,epsfig}
\usepackage{graphicx}
\allowdisplaybreaks
\begin{document}
\title{\bf Analysis of Bouncing Cosmology in Non-Riemannian Geometry}
\author{M. Sharif \thanks{msharif.math@pu.edu.pk}, M. Zeeshan Gul
\thanks{mzeeshangul.math@gmail.com}~ and~ Nusrat Fatima
\thanks{nusratfatimaliaqat@gmail.com}\\
Department of Mathematics and Statistics, The University of Lahore,\\
1-KM Defence Road Lahore-54000, Pakistan.}

\date{}
\maketitle

\begin{abstract}
The main objective of this manuscript is to investigate the bouncing
cosmology in the background of $f(\mathcal{Q})$ gravity, where
$\mathcal{Q}$ defines the non-metricity. For this purpose, we use
the reconstruction approach and consider a flat
Friedmann-Robertson-Walker spacetime with perfect matter
configuration. We examine how the first contracting phase gives the
expansion by using a temporal derivative of the scale factor, i.e.,
$\dot{a}<0$, $\dot{a}=0$ and $\dot{a}>0$ give contraction, bounce
point and expansion phases, respectively. Further, we use the order
reduction method to solve the modified field equations as these are
very difficult due to the presence of additional non-linear
expressions. It is analyzed that the original singularity of the
universe diminishes for the required bounce conditions. We conclude
that the acceleration occurs near the bouncing point and the
considered $f(\mathcal{Q})$ models are consistent with the current
cosmic accelerated expansion.
\end{abstract} {\bf Keywords:} Modified theories;
Singularities; Stability analysis.\\
{\bf PACS:} 04.20.Dw; 04.40.Dg; 04.20.Jb; 04.50.Kd.

\section{Introduction}

The general theory of relativity (GR), formulated by Einstein, is a
fundamental concept in physics that transformed our comprehension of
gravity and the structure of spacetime. It is a cornerstone of
modern physics and has been tested through observations and
experiments. However, this theory is based on geometric structures
in Riemann's metric space. Weyl \cite{1} developed a more general
geometrical structure that goes beyond Riemannian space and provides
a comprehensive explanation of gravitational fields and matter. His
objective was to unify gravitational and electromagnetic forces, not
all fundamental forces. The Levi-Civita connection is an essential
concept in Riemann metric space, which is used to compare vectors
based on their length. Weyl introduced a new type of connection that
does not consider the size of vectors during parallel transport. To
address the absence of information about vector's length, Weyl
introduced an additional connection known as the \emph{length
connection}. The length connection does not focus on the direction
of vector transport but instead on fixing or gauging the conformal
factor. Non-Riemannian geometries extend Riemannian geometry for
more general descriptions of spacetime curvature. These geometries
include torsion (twisting or rotation) or non-metricity (deviation
from metric compatibility). Weyl's theory incorporates the notion of
non-metricity with non-zero covariant derivative of the metric
tensor \cite{3}.

When we deviate from the assumption of a connection, it is possible
to develop alternative theories that are equivalent to GR. According
to the above discussion, there are two equivalent geometric
representations that can be used to represent GR, i.e., the
curvature representation disappears torsion and non-metricity
whereas the teleparallel representation disappears curvature and
non-metricity. However, one more comparable representation of the
fundamental geometric features is the non-metricity. The spacetime
is defined by the torsion when the curvature is set to zero, and the
resultant theory is known as the teleparallel equivalent to GR
(TEGR), also known as $f(\mathcal{T})$ theory, where $\mathcal{T}$
is the torsion scalar \cite{4}. This scenario considers a more
comprehensive structure known as the Weitzenb$\ddot{o}$ck connection
\cite{5}. By defining a non-metricity scalar $\mathcal{Q}$, the
corresponding action is found to be equivalent to that of TEGR. This
variant is termed as symmetric teleparallel equivalent to GR
(STEGR). The integral action for TEGR and STEGR is defined as $\int
d^{4}x\sqrt{-g} T $ and $\int d^{4}x\sqrt{-g}\mathcal{Q} $ ,
respectively \cite{6}. However, the extensions of GR, TEGR, and
STEGR are different from one another while $f(\mathcal{Q})$ is
equivalent to $f(\mathcal{T})$ and $f(R)$ theory leads to different
dynamics. Researchers have developed a range of extended
gravitational theories to address these challenges and delve deeper
into the mysteries of the cosmos \cite{6a}-\cite{6f}. Few studies
have been done in this direction so far, since it is innovative in
how it investigates some insights about the universe. Lu et al
\cite{7} studied some fascinating cosmic outcomes in STEGR. This
theory is appealing because it uses second order field equations,
which are simple to solve. Jimenez et al \cite{8} proposed the
$f(\mathcal{Q})$ theory which extends the action of STEGR to include
a basic function of non-metricity.

The analysis of bouncing cosmology in the framework of alternative
gravitational theories has gained a lot of attention in recent years
due to its fascinating characteristics. Cai et al \cite{1a} proposed
the detailed discussion on the matter bounce in $f(\mathcal{T})$
theory. Amani \cite{1b} studied the reconstruction of the bouncing
cosmos and investigated the stability of the cosmic models in $f(R)$
theory. Hohmann et al \cite{1c} analyzed the dynamical systems and
general characteristics of the cosmos in $f(R)$ theory. Shabani and
Ziaie \cite{1d} studied the bouncing behavior of the cosmos in
$f(R,T)$ theory, where $T$ is the trace of stress-energy tensor.
Sharif and Saba \cite{1e} studied the cosmography of generalized
ghost DE in $f(G)$ and $f(G,T)$ theories, where $G$ is the
Gauss-Bonnet invariant. Bhattacharjee and Sahoo \cite{1f} provided a
detailed study of a non-singular bounce in $f(R,T)$ theory. Bhardwaj
et al \cite{1g} introduced the cosmographic evolution of the closed
bouncing universe with a variable cosmological constant in the same
theory. Lazkoz et al \cite{2a} investigated the cosmological and
observational limitations of $f(\mathcal{Q})$ theory and showed that
the accelerating expansion is an intrinsic property of the geometry
of the universe. Mandal et al \cite{2d} examined the energy
conditions (ECs) and limited the model parameters to the current
values of cosmic parameters under the framework of $f(\mathcal{Q})$
theory to verify the viability of their cosmic models. Mandal et al
\cite{2c} performed a detailed cosmographic analysis in the same
theory. Bajardi et al \cite{2e} studied the Hamiltonian and ADM
formalism to determine the wave function of the universe within this
framework. Mandal et al \cite{2f} explored the bouncing scenarios
with two different Lagrangian forms of $f(\mathcal{Q})$ using
perturbation technique and concluded that the value of the perturbed
term is very high at the bouncing point and later converges towards
zero.

In this paper, we study bouncing cosmology through the order
reduction method of the field equations in $f(\mathcal{Q})$ gravity.
This paper is structured as follows. We obtain the solution of the
gravitational field equations by using the parametrization approach
on the Hubble parameter. We propose detailed information on the
cosmic models and examine the bouncing criteria in section
\textbf{2}. We use a reconstruction approach to investigate the
cosmological behavior of the universe by using two cosmological
models in section \textbf{3}. The rebuilt redshift model is used to
specify effective energy density in section \textbf{4}. In section
\textbf{5}, we summarize our results.

\section{Modified Symmetric Teleparallel Gravity}

The integral action for $f(Q)$ theory is defined as \cite{2c}
\begin{equation}\label{1}
\mathcal{S}=\int\left(\frac{1}{2\kappa}f(\mathcal{Q})+L_{m}\right){\sqrt{-g}}
d^4x,
\end{equation}
where $\kappa$ is the coupling constant and $L_{m}$ represents the
matter Lagrangian density. The corresponding field equations are
\begin{equation}\label{2}
-\frac{2}{\sqrt{-g}}\nabla_{\phi}(f_\mathcal{Q}{\sqrt{-g}}P^{\phi}_{\mu\nu})
-\frac{1}{2}f(\mathcal{Q})g_{\mu\nu}-f_\mathcal{Q}(P_{\mu\phi\tau}\mathcal{Q}
_{\nu}^{~~\phi\tau}-2\mathcal{Q}^{\phi\tau}_{~~\mu}P_{\phi\tau\nu})={\kappa}T
_{\mu\nu}.
\end{equation}
Here $f_{\mathcal{Q}}=\frac{\partial f}{\partial\mathcal{Q}}$ and
$\nabla_{\phi}$ denotes the covariant derivative. Using
superpotential, the non-metricity scalar is defined as
\begin{equation}\label{3}
\mathcal{Q}=-\mathcal{Q}_{\phi\mu\nu}P^{\phi\mu\nu}
=-\frac{1}{4}\big[-\mathcal{Q}^{\phi\mu\nu}\mathcal{Q}_{\phi\mu\nu}
+2\mathcal{Q}^{\phi\mu\nu}\mathcal{Q}_{\nu\phi\mu}-2\mathcal{Q}^{\phi}
\tilde{\mathcal{Q}}_{\phi}+\mathcal{Q}^{\phi}\mathcal{Q}_{\phi}\big],
\end{equation}
where
\begin{equation}\label{4}
\mathcal{Q}_{\phi\mu\nu}=\nabla_{\phi}g_{\mu\nu} \neq 0, \quad
\mathcal{Q}_{\phi}=\mathcal{Q}_{\phi~~\mu}^{~~\mu},\quad
\tilde{\mathcal{Q}}_{\phi}=\mathcal{Q}^{\mu}_{~~\phi\mu}.
\end{equation}
The calculation of the above relation is shown in Appendix
\textbf{A} and the explicit formulation of $\delta \mathcal{Q}$ is
given in Appendix \textbf{B}. Furthermore, we can write the
superpotential as
\begin{equation}\label{5}
P^{\phi}_{~\mu\nu}=-\frac{1}{2}L^{\phi}_{~\mu\nu}+
\frac{1}{4}(\mathcal{Q}^{\phi}-\tilde{\mathcal{Q}}^{\phi})g_{\mu\nu}
-\frac{1}{4}\delta^{\phi}~_{(\mu}\mathcal{Q}_{\nu)}.
\end{equation}
The isotropic matter configuration is given as
\begin{equation}\label{6}
T_{\mu\nu}=(p+\rho)u_{\mu}u_{\nu}-pg_{\mu\nu},
\end{equation}
where $\rho,~p$ and $u_{\mu}$ are the energy density, pressure and
four-velocity of the fluid, respectively.

We consider flat FRW line element to examine the mysterious features
of the cosmos as
\begin{equation}\label{7}
ds^2=dt^{2}-(dx^{2}+dy^{2}+dz^{2})a^{2}(t),
\end{equation}
where the scale factor is denoted by $a$. Equations (\ref{3}) and
(\ref{7}) enable us to write the trace of  the non-metricity tensor
(details are given in Appendix $\textbf{C}$) as
\begin{equation}\label{8}
\mathcal{Q}=-6H^{2},
\end{equation}
where $H=\frac{\dot{a}^{2}}{a^{2}}$ is the Hubble parameter and dot
means derivative with respect to time $t$. The resulting field
equations are
\begin{eqnarray}\label{9}
\rho&=&-6H^{2}f_\mathcal{Q}-\frac{1}{2}f(\mathcal{Q}),
\\\label{10}
p&=&\frac{1}{2}f(\mathcal{Q})
+2f_{\mathcal{Q}}\dot{H}+2f_{\mathcal{Q}\mathcal{Q}}H
+6f_{\mathcal{Q}}H^{2}.
\end{eqnarray}
These field equations are helpful to examine the ambiguous
characteristics of the universe. Harko et al \cite{27} studied
power-law and an exponential form of $f(\mathcal{Q})$ to study the
mysterious universe. We consider the following two models of
$f(\mathcal{Q})$ gravity.

\subsection*{Model 1}

The linear $f(\mathcal{Q})$ model is given as
\begin{equation}\label{11}
f(\mathcal{Q})=\mathcal{Q}\psi,
\end{equation}
where $\psi$ is a non-zero arbitrary constant. Inserting this model
in Eqs.(\ref{9}) and (\ref{10}), we obtain
\begin{eqnarray}\label{12}
\rho&=&-3{\psi}H^{2},
\\\label{13}
p&=&\psi(3H^2+2\dot{H}).
\end{eqnarray}
This model explains the cosmic growth and evolution. We compare the
relevant outcomes with the standard CDM cosmology and theoretical
predictions of the $f(\mathcal{Q})$ theory. It follows that the
universe started in a decelerating phase for the specific range of
cosmological parameters and can reach to the de Sitter phase in the
late-time.

\subsection*{Model 2}

Here, we assume a non-linear model as
\begin{equation}\label{14}
f(\mathcal{Q})=\psi\mathcal{Q}^{m+1}.
\end{equation}
The corresponding field equations (\ref{9}) and (\ref{10}) are given
by
\begin{eqnarray}\label{15}
\rho&=&-\frac{1}{2}[\psi(-6H^{2})^{m+1}]-6H^{2}[\psi
(m+1)(-6H^{2})^{m}],
\\\nonumber
p&=&\frac{1}{2}[\psi(-6H^{2})^{m+1}]+[2\psi(m+1)(-6H^{2})^m](\dot{H}+3H^{2})
\\\label{16}
&+&2[\psi m(m+1)(-6H^{2})^{m-1}H.
\end{eqnarray}
This model determines the change from deceleration to acceleration
phase. By a suitable choice of the model parameters, the accelerated
expansion can be achieved easily.

\section{Analysis of Bouncing Cosmology}

In cosmology, bouncing solutions are significant because they
resolve the initial singularity which describes the universe's
expansion. Accordingly, the big bang and big crunch phases take
place one after the other regularly. In a cyclic universe scenario,
the cosmos gradually moves from an earlier contraction phase to an
expansion phase without experiencing a singularity. The cosmic
bounce may be viewed as a rhythmic or periodic cosmos in which the
collapse of one phase leads to the occurrence of another
cosmological event. Cai et al \cite{28} studied the symmetric bounce
model to create a non-singular bouncing cosmos after a contraction
phase. This bounce prevents the basic initial singularity issue when
combines with other cosmic occurences \cite{29}. The symmetric
bouncing cosmology avoids a singularity similar to the big bang. The
following concepts build a realistic bouncing behavior of the
universe.
\begin{itemize}
\item
The decreasing behavior of the scale factor represents that the
cosmos is in the contracting phase while the increasing behavior
determines the cosmic expansion era.
\item
The temporal derivative of the scale factor must be zero near the
bouncing point. The scale factor must be minimum near the bouncing
spot for the non-singular bouncing model.
\item
At the cosmic bounce, the cosmos contracts, expands and reaches the
bouncing point when $H<0$, $H>0$ and $H=0$, respectively.
\item
The equation of state (EoS) parameter must pass through the phantom
divide line at the bouncing point.
\end{itemize}

\subsection{Reconstruction Approach}

In this section, we examine whether it is possible to obtain an
appropriate gravitational Lagrangian that can accurately reproduce
the cosmic growth defined by different cosmological models. This
technique allows the gravitational Lagrangian to be solved by a
selected cosmology, which can be established analytically using the
form $a$, or $H$,  or cosmological measurements. Further, to analyze
the cosmological changes, we take the Hubble parameter \cite{26} as
\begin{equation}\label{17}
H=\gamma t^{n}+\beta t+\alpha,
\end{equation}
where $n,~\alpha,~\beta$ and $\gamma$ are non-zero constants. This
parametrization aims to recreate a bouncing scenario with cosmic
acceleration, contraction before the bounce, and rapid expansion
after the bounce.
\begin{figure}
\epsfig{file=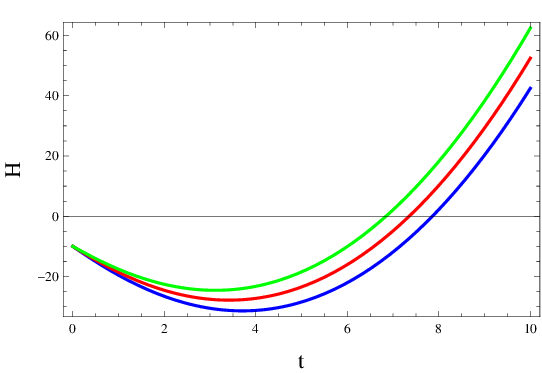,width=.5\linewidth}
\epsfig{file=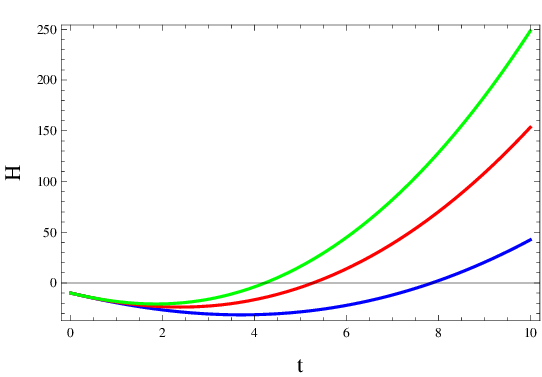,width=.5\linewidth}\caption{Evolution of Hubble
constant for various values of $\beta$ and $\gamma$.}
\end{figure}

Figure \textbf{1} represents the bouncing location for $\alpha=-10$,
$n=2.2$ and various values of $\beta,~\gamma$ corresponding to the
cosmic time. This implies that the cosmos changes from contracting
to expanding phases. We observe that the bouncing points occur at
$t=6.5$, $7.0$ and $7.9$ when $\gamma=1$ and $\beta=-8.6$, $-9.6$
and $-10.6$, respectively. According to the right graph, the
bouncing spots are at $t=4.2$, $5.2$ and $7.9$ for $\beta=-10.6$ and
$\gamma=2.3$, $1.7$ and $1$, respectively. We examine the bouncing
point $t=7.9$ for $\beta=-10.6$ and $\gamma=1$. Tables \textbf{1}
and \textbf{2} show the nature of the Hubble parameter for various
values of $\gamma$ and $\beta$ as well as fixed values of $n$ and
$\alpha$. Using Eq.(\ref{17}), the scale factor becomes
\begin{table}\caption{\textbf{Restrictions on model parameters with $\beta$ variation}}
\begin{center}
\begin{tabular}{|c|c|c|c|}
\hline $\beta$ & $\gamma$ & Time Interval & Behavior of $H$
\\
\hline -8.6   & 1      & $0<t<6.5$   & Contraction
\\
\hline -9.6   & 1      & $0<t<7.0$   & Contraction
\\
\hline -10.6    & 1      & $0<t<7.9$   & Contraction
\\
\hline -8.6   & 1   & $6.5<t<\infty$  & Expansion
\\
\hline -9.6   & 1    & $70<t<\infty$ & Expansion
\\
\hline -10.6    & 1    & $7.9<t<\infty$ & Expansion
\\
\hline
\end{tabular}
\end{center}
\end{table}
\begin{equation}\label{18}
a(t)=\eta e^{\frac{\gamma t^{n+1}}{n+1}+\frac{\beta t^{2}}{2}+\alpha
t},
\end{equation}
where the integration constant is denoted by $\eta$. In the left
panel of Figure $\textbf{2}$, we can observe that the bounce is at
$t=7.9$. Also, the Hubble parameter is negative before the bounce
and positive after the bounce with $\dot{H}>0$ at the bouncing
epoch. In the right panel of Figure $\textbf{2}$, the behavior of
the scale factor shows that the universe shifts from the contraction
phase $(t<7.9)$ to an expansion phase $(t>7.9)$. During the
bouncing, the scale factor does not appear at $t=7.9$.
\begin{table}\caption{\textbf{Restrictions on model parameters with $\gamma$ variation}}
\begin{center}
\begin{tabular}{|c|c|c|c|}
\hline $\beta$ & $\gamma$ & Time Interval & Behavior of $H$
\\
\hline -10.6   & 2.3       & $0<t<4.2$   & Contraction
\\
\hline -10.6   & 1.7      & $0<t<5.2$   & Contraction
\\
\hline -10.6   & 1      & $0<t<7.9$   & Contraction
\\
\hline -10.6   & 2.3      & $4.2<t<\infty$  & Expansion
\\
\hline -10.6   & 1.7      & $5.2<t<\infty$ & Expansion
\\
\hline -10.6   & 1      & $7.9<t<\infty$ & Expansion
\\
\hline
\end{tabular}
\end{center}
\end{table}
\begin{figure}
\epsfig{file=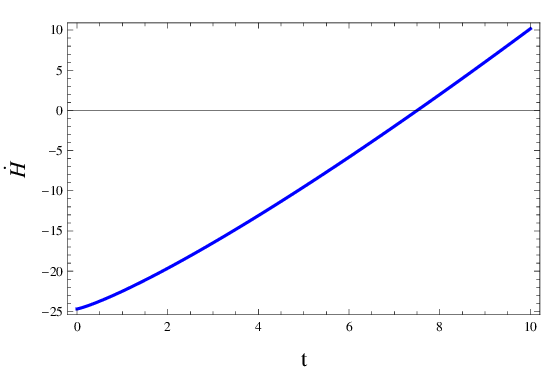,width=.5\linewidth}
\epsfig{file=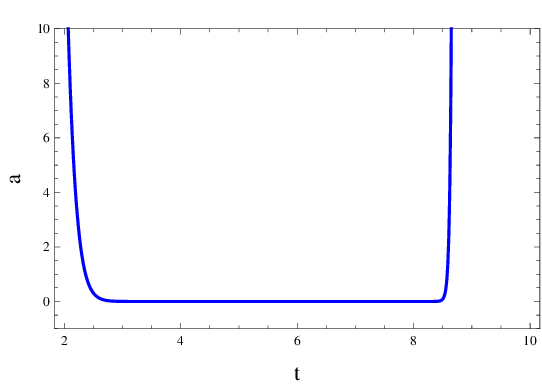,width=.5\linewidth}\caption{Behavior of temporal
derivative of the Hubble parameter and scale factor.}
\end{figure}

The deceleration parameter $(q)$ is another important cosmic
quantity that represents the increasing/decreasing behavior of
cosmic evolution. When $q>0$, the cosmos is said to be decelerating
while accelerating for $q<0$. This is expressed as
\begin{equation}\label{19}
q=-\frac{a\ddot{a}}{\dot{a}^{2}}=-1-\frac{\dot{H}}{H^{2}}.
\end{equation}
Using Eqs.(\ref{17}) and (\ref{19}), we have
\begin{equation}\label{20}
q=-1-\frac{ n \gamma t^{n-1} + \beta }{(\gamma t^{n} + \beta t
+\alpha)^2}.
\end{equation}
The graphical analysis of $q$ is shown in Figure \textbf{3}. The
deceleration parameter is $-1$ before and after the bouncing point,
indicating a negative value for both the contraction and expansion.
From Eq.(\ref{19}), the deceleration parameter $q<0$ remains
constant for all values of $t$, whereas acceleration is maximum
around the bouncing point $t\approx7.9$ (see Figure \textbf{3}).
Table \textbf{3} indicates that the behavior of deceleration
parameter is symmetric to the Hubble parameter.
\begin{figure}\center
\epsfig{file=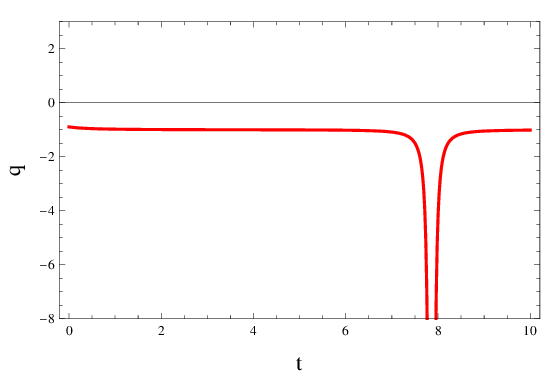,width=.5\linewidth}\caption{Plot of deceleration
constant versus cosmic time.}
\end{figure}
\begin{table}\caption{\textbf{Evolution of the universe for $\alpha=-10,~\beta=-10.6$
and $\gamma=1$}}
\begin{center}
\begin{tabular}{|c|c|c|c|}
\hline $q$ & $H$  & Time  & Behavior of the universe
\\
\hline $<0$   & $<0$      & $0<t<7.9$         & $q{\oplus}H$
\\
\hline $<0$   & $=0$      & $t\approx{7.9}$ &$q_{max}{\oplus}H$
\\
\hline $<0$   & $>0$      & $7.9<t<\infty$    & $q{\oplus}H$
\\
\hline
\end{tabular}
\end{center}
\end{table}

The corresponding Eqs.(\ref{12}) and (\ref{13}) for model 1 are
\begin{eqnarray}\label{21}
\rho&=&-3{\psi}{(\gamma t^{n}+\beta t+\alpha)^2},
\\\label{22}
p&=&\psi[3{(\gamma t^{n}+\beta t+\alpha)^2}+2(n \gamma
t^{n-1}+\beta)].
\end{eqnarray}
Equations (\ref{15}) and (\ref{16}) with respect to model 2 are
\begin{eqnarray}\nonumber
\rho&=&-\frac{1}{2}[\psi(-6(\gamma t^{n}+\beta t+\alpha)^{2})^{m+1}]
\\\label{23}
&-&6(\gamma t^{n}+\beta t+\alpha)^{2}[\psi (m+1)(-6(\gamma
t^{n}+\beta t+\alpha)^{2})^{m}].
\\\nonumber
p&=&\frac{1}{2}[\psi(-6(\gamma t^{n}+\beta t+\alpha)^{2})^{m+1}]
\\\nonumber
&+&[2\psi(m+1)(-6{(\gamma t^{n}+\beta t +\alpha)^2})^m][ (n\gamma
t^{n-1}+\beta)\\\nonumber &+&3{(\gamma t^{n}+\beta t+\alpha)^2}]
\\\label{24}
&+&2\psi m(m+1)[-6{(\gamma t^{n}+\beta t+\alpha)^2}]^{m-1}(\gamma
t^{n}+\beta t+\alpha).
\end{eqnarray}
The EoS parameter in Figure \textbf{4} is non-singular at the
bouncing epoch and develops contraction in the vicinity of the
bounce. In this instance, the EoS parameter is symmetric around the
bouncing epoch. It determines how the EoS parameter passes the
phantom divide line $\omega=-1$. We plot the graphs for
$\beta=-10.6$ and $\gamma=1$ in Figure \textbf{5}, which
demonstrates that the NEC is violated when the bouncing requirements
are met in the range $0\leq t \leq7.9$.

\subsection{Energy Conditions}
\begin{figure}
\epsfig{file=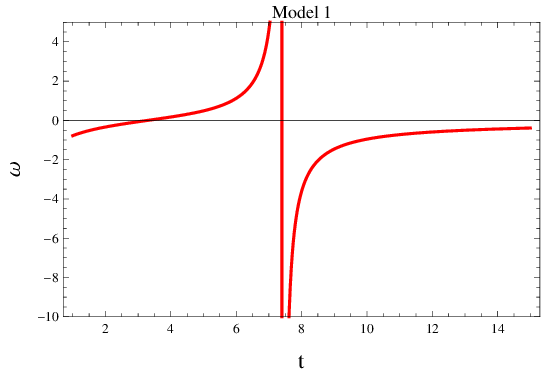,width=.5\linewidth}
\epsfig{file=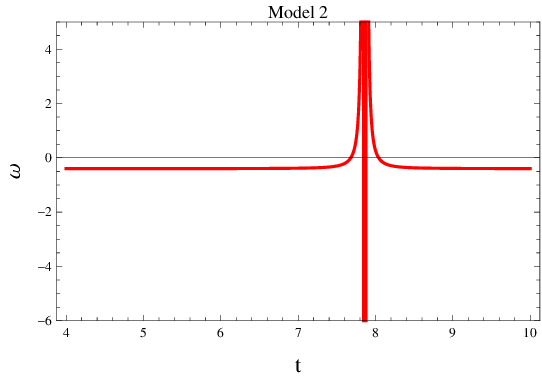,width=.5\linewidth}\caption{Graphs of EoS
parameter.}
\end{figure}

To comprehend the geodesic of the universe under various energy
conditions (ECs) is another important investigation. The conditions
used to examine the behavior of stress-energy tensor in the
existence of matter are known as energy constraints. There are four
types of ECs named as null (NEC), weak (WEC), strong (SEC) and
dominant energy condition. These requirements possess limitations on
certain linear combinations of energy density and pressure. This
leads to the conclusion that gravity always shows an attractive
force and therefore energy density cannot be negative. These are
classified as \cite{30}
\begin{figure}
\epsfig{file=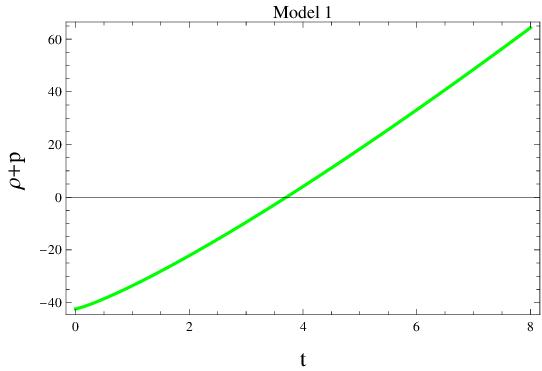,width=.5\linewidth}
\epsfig{file=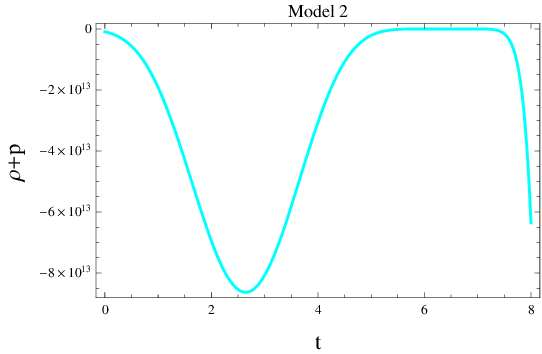,width=.5\linewidth}\caption{Evolution of the
NEC.}
\end{figure}
\begin{itemize}
\item
DEC: $\rho\geq|p|$,
\item
NEC: $\rho+p\geq0$,
\item
SEC: $\rho+p\geq0$, $(n-3)\rho+(n-1)p\geq0$,
\item
WEC: $\rho\geq0$, $\rho+p\geq0$.
\end{itemize}
Among all the ECs, the NEC is significant as it must be satisfied
for any stable system. The violation of NEC ensures the violation of
all ECs as these conditions depend on NEC. It is well-known that a
successful non-singular isotropic bounce needs to violate the NEC.
The well-known Hawking-Penrose singularity invokes the SEC and its
violation leads the observed accelerated expansion \cite{31}. If the
NEC violates then WEC and DEC cannot be met with SEC. Figures
\textbf{6} and \textbf{7} show the graphical analysis of ECs
implying that there are no singularities around the bouncing epoch.
The ECs $\rho+p$ and $\rho+3p$ are negative close to the bounce
point by the expected matter bounce scenario and as a result, the
energy criteria violates.
\begin{figure}
\epsfig{file=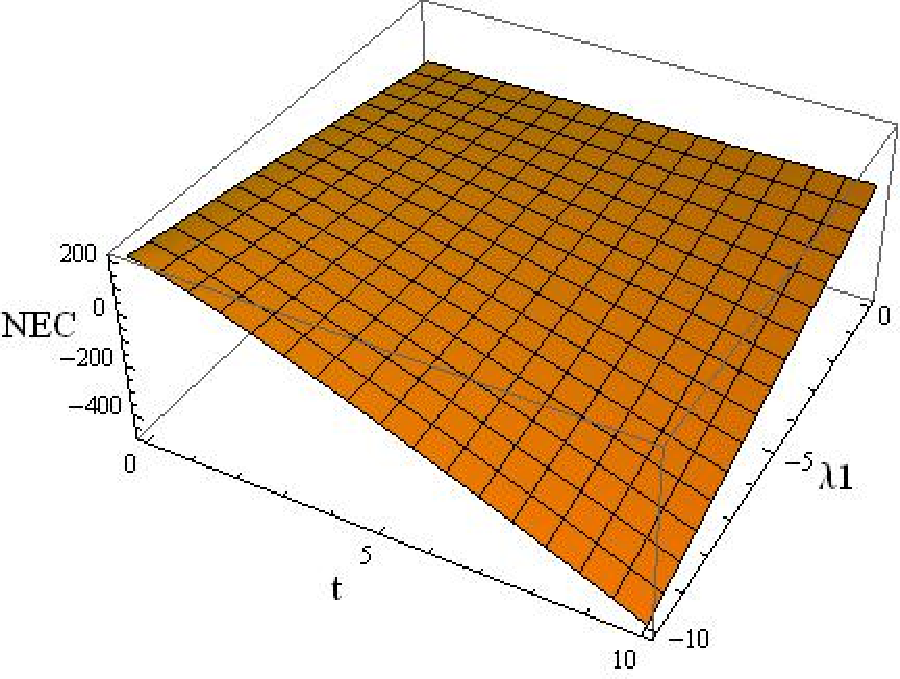,width=.5\linewidth}
\epsfig{file=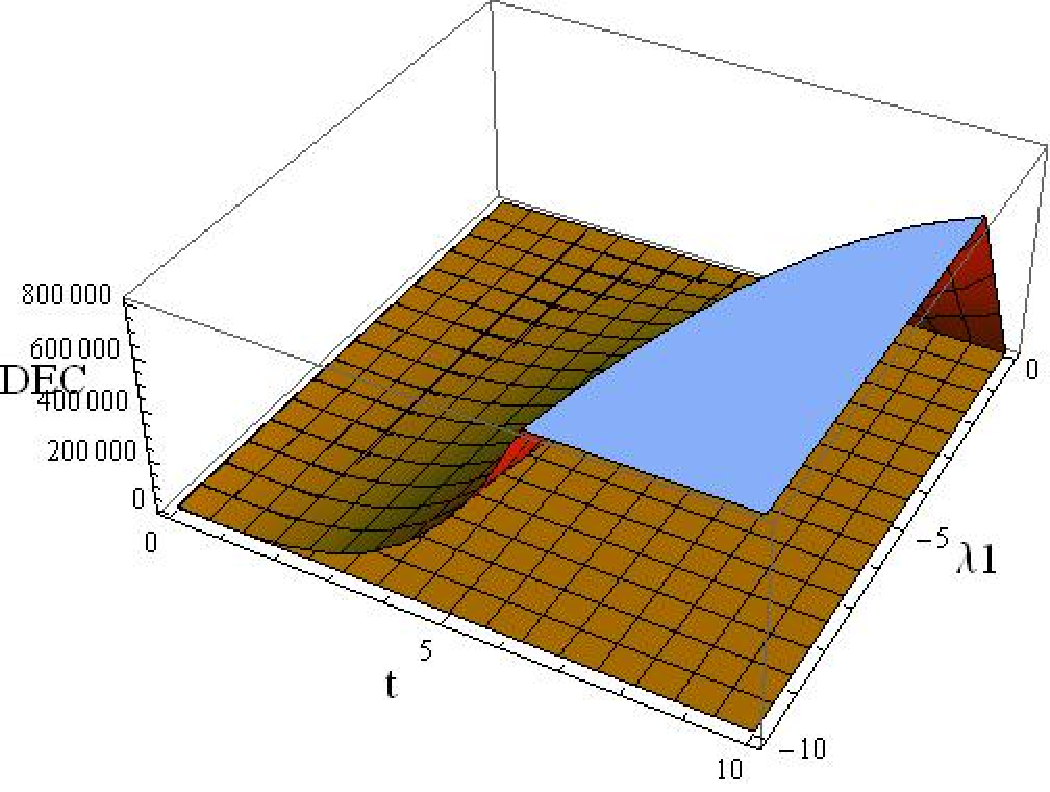,width=.5\linewidth}\center
\epsfig{file=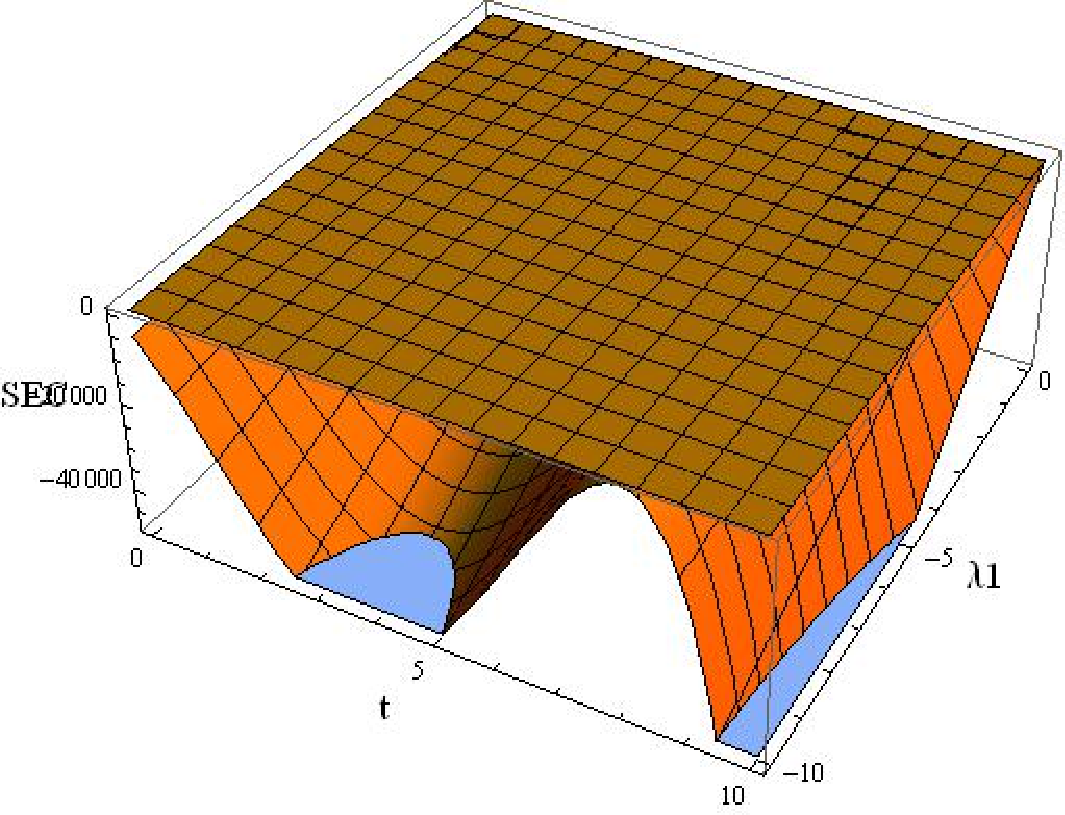,width=.5\linewidth}\caption{Behavior of energy
conditions for $n=2.2$ for model $1$.}
\end{figure}
\begin{figure}
\epsfig{file=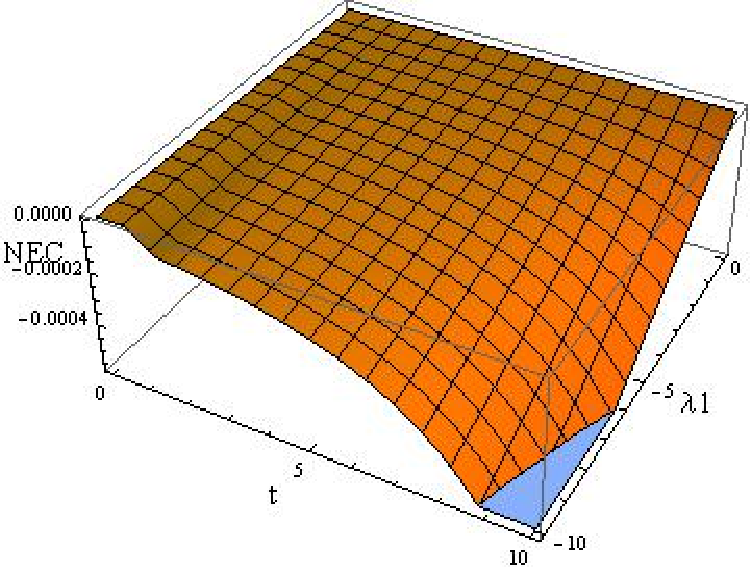,width=.5\linewidth}
\epsfig{file=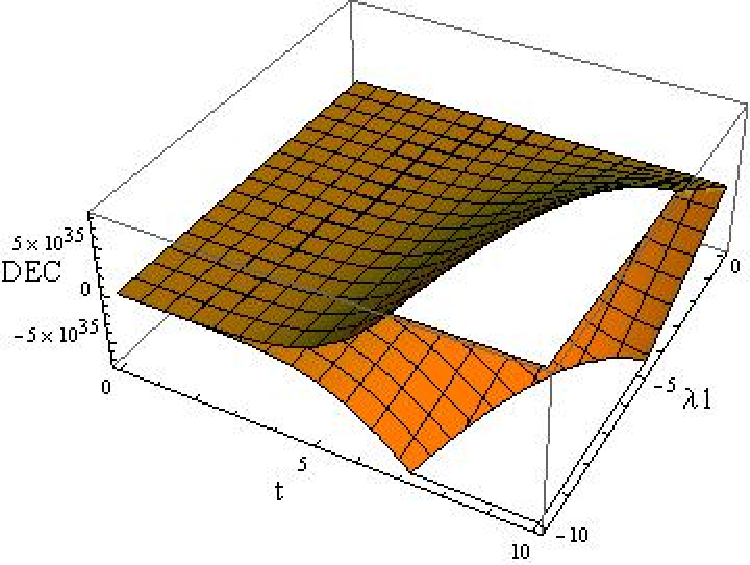,width=.5\linewidth}\center
\epsfig{file=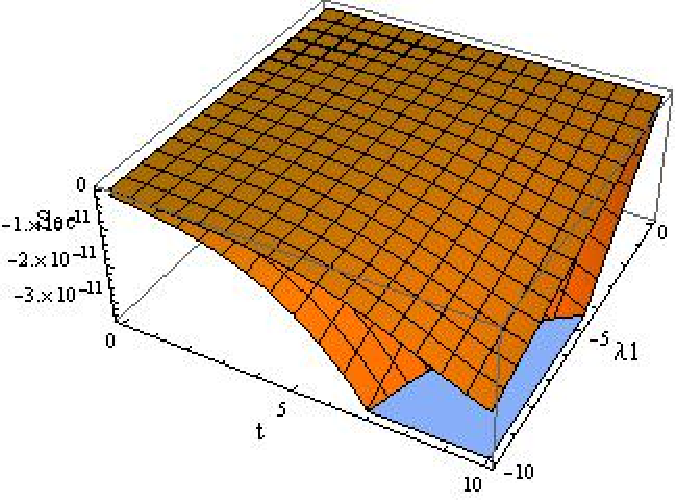,width=.5\linewidth}\caption{Behavior of energy
conditions for $n=2.2$ for model $2$.}
\end{figure}

\section{Reparametrization By Red-Shift}

Here, we analyze the model of the redshift parameter. For this
purpose, we consider $z+1=\frac{a_{0}}{a}$, where $a_{0}$ represents
the value in the current time. Moreover, we take dimensionless
parameter as \cite{32}
\begin{equation}\label{25}
r(z)=\frac{H^{2}(z)}{H_{0}},
\end{equation}
where $H_{0}=71\pm3 km s^{-1} Mpc^{-1}$. The following differential
form represents the relation between redshift and cosmic time as
\begin{equation}\label{26}
\frac{d}{dt}=\frac{da}{dt}\frac{dz}{da}\frac{d}{dz}=-H(z+1)\frac{d}{dz}.
\end{equation}
The expansion rate of the universe is measured by the Redshift
observation. In 1962, Sandage \cite{33} proposed to directly measure
the variation of the redshift of distant sources. Here, we take two
models and reconstruct them in terms of redshift. We focus on the
late-time expansion, so that we can ignore radiation and consider
the entire contribution due to pressureless matter. Using
Eqs.(\ref{8}) and (\ref{9}), the non-metricity and the effective
energy density turn out to be
\begin{figure}
\epsfig{file=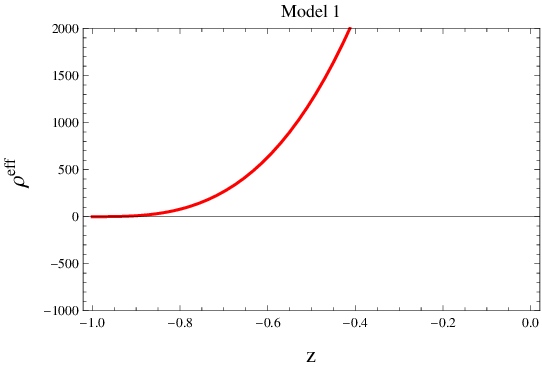,width=.5\linewidth}
\epsfig{file=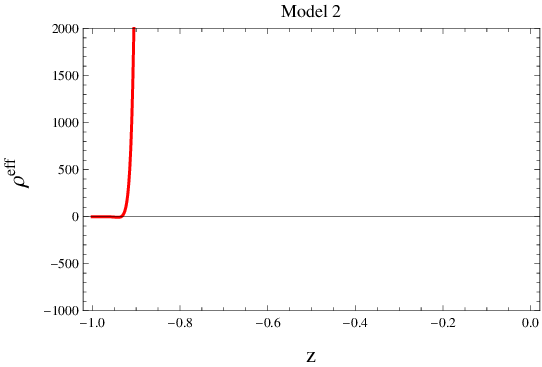,width=.5\linewidth}\caption{Graphs of matter
variable for models 1 and 2.}
\end{figure}
\begin{eqnarray}\label{27}
\mathcal{Q}&=&-6H^{2}_{0}r(z),
\\\label{28}
\rho^{eff}&=&-6H_{0}^{2}r(z)f^{\prime}(z)-\frac{1}{2}f(z).
\end{eqnarray}
To recreate the models, we add the $r(z)$ feature as \cite{34}
\begin{equation}\label{29}
r(z)=\Omega_{m0}(z+1)^{3},
\end{equation}
where $\Omega_{m0}=0.3$ \cite{35}. The cosmic parameters in the form
of redshift are obtained by substituting Eq.(\ref{29}) into
(\ref{28}) for models \textbf{1} and \textbf{2} and the results are
displayed in Figure \textbf{8}. The behavior of effective energy
density in terms of redshift determines that  $\rho^{eff}$ is
positive, which supports the cosmic accelerated expansion and thus
correlates with the observational results. The energy density in
terms of redshift is calculated through the fixed values of model
parameters $0<\psi<1$ and $m=3$.

\section{Conclusions}

In this paper, we have analyzed the bouncing cosmology in
$f(\mathcal{Q})$ gravity using the order reduction approach to find
bouncing solutions. Because modified theories of gravity contain
extra degrees of freedom, solving their field equations is
challenging. The order reduction approach is therefore helpful in
the study of the cosmic evolution. We have considered the STEGR
action as a non-metric tool and applied the reconstruction as well
as redshift approach to examine the cosmic expansion in this theory
of gravity. The effective energy density function for the modified
gravity is constructed. The bouncing behavior of two distinct models
of $f(\mathcal{Q})$ gravity has been explored in terms of cosmic
time. We have obtained the bouncing state at the bounce position and
demonstrated the associated cosmological parameters. The cosmic
parameters and the effective energy density have been stated in
terms of redshift parameter. We have then reconstructed the models
with a redshift component using the $r(z)$ function parametrization.
The main results are given as follows.
\begin{itemize}
\item
It is observed that the bouncing points occur for decreasing values
of $\beta$ and $\gamma$ (Figure \textbf{1}).
\item
Tables \textbf{1} and \textbf{2} illustrate the range of $\beta$ and
$\gamma$ that may be used to achieve the bouncing behavior between a
contraction phase $(H<0)$ and an expansion phase $(H>0)$.
\item
The temporal derivative of Hubble parameter represents that the
universe shifts from contraction to expansion phase around the
bouncing point $t=7.9$ (left plot of Figure \textbf{2}). The right
graph determines that the scale factor decreases during the
contraction phase $(t<7.9)$ and increases during the expansion phase
$(t>7.9)$. During the bouncing, the scale factor does not appear at
$t=7.9$.
\item
It is found that the universe is accelerating and attains its
maximum value at the bouncing spot $t=7.9$, when $q<0$ (Figure
\textbf{3}).
\item
The EoS parameter changes from $\omega<-1$ to $\omega=-1$ near the
bouncing point $t=7.9$ (Figure \textbf{4}).
\item
The NEC is violated when the bouncing requirements are satisfied
(Figures \textbf{5}-\textbf{7}).
\item
The behavior of $\rho^{eff}$ (in terms of redshift) indicates that
$\rho^{eff}$ varies positively, which confirms the rapid expansion
of the cosmos (Figure \textbf{8}).
\end{itemize}

We  have found that the cosmos experiences the bounce period in the
future, indicating that the violation of NEC in the past will become
obvious \cite{36}. We have tested theoretical consistency of the
novel STEGR theory. This method also inspires and promotes research
into other $f(\mathcal{Q})$ types of theories and extensions. Our
results indicate that the universe is undergoing decelerated
expansion and then reaches to accelerated expansion phase. This is
consistent with the standard CDM model \cite{2d} and de Sitter type
expansion. We have compared our (ECs) with the CDM model. In CDM
model, all ECs are satisfied except SEC. This behavior is not
compatible with our proposed models as we have found that NEC is
violated when the bouncing requirements are satisfied. In our case,
the EoS parameter for the two $f(\mathcal{Q})$ models is consistent
with the current phase of rapid expansion, presenting values nearly
equal to $-1$. This behavior is satisfied with the CDM description
as well as current accelerated expansion \cite{37}. We can conclude
this work provides some fundamental conceptual tools for the
detailed examination of the geometric features of gravity and its
consequences in the background of cosmology.

Bouncing cosmology proposes an alternative to the big bang model by
suggesting that the universe undergoes cycles of contraction and
expansion with a bounce occurring at the transition between these
phases. This concept addresses some of the shortcomings of the big
bang theory such as the initial singularity and the horizon problem.
On the other hand, $f(\mathcal{Q})$ gravity is a modified gravity
theory where the action is described by a function of non metricity.
This modification allows for a departure from GR and provides a
framework to incorporate additional gravitational effects into
cosmological models. When combining bouncing cosmology with
$f(\mathcal{Q})$ gravity, we obtain a novel approach to understand
the evolution of the universe. This combination offers several
predictions and testable features as
\begin{itemize}
\item
Bounce Dynamics
\\\\
One of the primary predictions of bouncing cosmology is the
existence of a non-singular bounce, where the universe transitions
from a contracting phase to an expanding phase without encountering
a singularity. In the framework of $f(\mathcal{Q})$ gravity, the
specific form of the function influences the dynamics of the bounce
and the subsequent evolution of the universe.
\item
Primordial Perturbations
\\\\
Bouncing cosmology predicts distinctive features in the primordial
perturbations of the cosmic microwave background radiation. These
features can be compared with observational data from experiments
such as the Planck satellite to test the viability of the model.
\item
Cosmological Parameters
\\\\
The combination of bouncing cosmology with $f(\mathcal{Q})$ gravity
leads to modifications in the evolution of cosmological parameters
such as the expansion rate of the universe. Observational
constraints on these parameters can provide insights into the
validity of the model.
\item
Cosmic Microwave Background Anisotropy
\\\\
Bouncing cosmology in $f(\mathcal{Q})$ gravity predicts specific
patterns of anisotropy in the cosmic microwave background which
differ from those predicted by cosmological models.
\end{itemize}
These features in the framework of bouncing cosmology in
$f(\mathcal{Q})$ gravity assess the viability of this approach and
give deeper insights into the fundamental properties of the
universe.

\section*{Appendix A: Computation of
$\mathcal{Q}=-\mathcal{Q}_{\phi\mu\nu}P^{\phi\mu\nu}$}
\renewcommand{\theequation}{A\arabic{equation}}
\setcounter{equation}{0}

Using Eqs.(\ref{2}) and (\ref{3}), we have
\begin{eqnarray}\label{A1}
\mathcal{Q}&=&-g^{\mu\nu}\big(L^{\phi}_{~\alpha\mu}L^{\alpha}
_{~\nu\phi}-L^{\phi}_{~\alpha\phi}L^{\alpha}_{~\mu\nu}\big),
\\\label{A2}
L^{\phi}_{~\alpha\mu}&=&-\frac{1}{2}g^{\phi\lambda}\big(\mathcal{Q}
_{\mu\alpha\lambda}+\mathcal{Q}_{\alpha\lambda\mu}
-\mathcal{Q}_{lambda\mu\alpha}\big),
\\\label{A3}
L^{\alpha}_{~\nu\phi}&=&-\frac{1}{2}g^{\alpha\beta}\big(\mathcal{Q}
_{\phi\nu\beta}+\mathcal{Q}_{\nu\beta\phi}-\mathcal{Q}_{\beta\phi\nu}\big),
\\\label{A4}
L^{\phi}_{~\alpha\phi}&=&-\frac{1}{2}g^{\phi\lambda}\big(\mathcal{Q}_{\phi\alpha\lambda}
+\mathcal{Q}_{\alpha\lambda\phi}-\mathcal{Q}_{\lambda\phi\alpha}\big),
\\\label{A5}
L^{\alpha}_{~\mu\nu}&=&-\frac{1}{2}g^{\alpha\beta}\big(\mathcal{Q}_{\nu\mu\beta}
+\mathcal{Q}_{\mu\beta\nu}-\mathcal{Q}_{\beta\nu\mu}\big).
\end{eqnarray}
Therefore, we obtain
\begin{eqnarray}\label{A6}
-g^{\mu\nu}L^{\phi}_{~\alpha\mu}L^{\alpha}_{~\nu\phi}
&=&-\frac{1}{4}\big(2\mathcal{Q}^{\phi\nu\beta}\mathcal{Q}_{\beta\phi\nu}
-\mathcal{Q}^{\phi\nu\beta} \mathcal{Q}_{\phi\nu\beta}\big),
\\\label{A7}
g^{\mu\nu}L^{\phi}_{~\alpha\phi}L^{\alpha}_{~\mu\nu}
&=&\frac{1}{4}g^{\mu\nu}g^{\alpha\beta}\mathcal{Q}_{\alpha}\big(\mathcal{Q}
_{\nu\mu\beta}+\mathcal{Q}_{\mu\beta\nu}-
\mathcal{Q}_{\beta\nu\mu}\big),
\\\nonumber
\mathcal{Q}&=&-\frac{1}{4}\big(-\mathcal{Q}^{\phi\nu\beta}\mathcal{Q}_{\phi\nu\beta}
+2\mathcal{Q}^{\phi\nu\beta}\mathcal{Q}_{\beta\phi\nu}
\\\label{A8}
&-&2\mathcal{Q}^{\beta}\tilde{\mathcal{Q}}_{\beta}
+2\mathcal{Q}^{\beta}\mathcal{Q}_{\beta}\big).
\end{eqnarray}
By considering Eq.(\ref{5}), we get
\begin{eqnarray}\nonumber
P^{\phi\mu\nu}&=&\frac{1}{4}\bigg[-\mathcal{Q}^{\phi\mu\nu}
+\mathcal{Q}^{\mu\phi\nu}+\mathcal{Q}^{\nu\phi\mu}
+\mathcal{Q}^{\phi}g^{\mu\nu}-\tilde{\mathcal{Q}}^{\phi}g^{\mu\nu}
\\\label{A9}
&-&\frac{1}{2}(g^{\phi\mu}\mathcal{Q}^{\nu}+g^{\phi\nu}\mathcal{Q}^{\mu})\bigg].
\\\nonumber
-\mathcal{Q}_{\phi\mu\nu}P^{\phi\mu\nu}&=&-\frac{1}{4}
\big(-\mathcal{Q}^{\phi\mu\nu}\mathcal{Q}_{\phi\mu\nu}
+2\mathcal{Q}_{\phi\mu\nu}\mathcal{Q}^{\mu\phi\nu}
+\mathcal{Q}^{\phi}\mathcal{Q}_{\phi}-2\mathcal{Q}_{\phi}
\tilde{\mathcal{Q}}^{\phi}\big)
\\\label{A10}
&=&\mathcal{Q}.
\end{eqnarray}
Applying the relation
$\mathcal{Q}_{\phi\mu\nu}\mathcal{Q}^{\mu\phi\nu}
=\mathcal{Q}_{\phi\mu\nu}\mathcal{Q}^{\nu\phi\mu}$, we are able to
get the above result.

\section*{Appendix B: Calculation of $\delta\mathcal{Q}$}
\renewcommand{\theequation}{B\arabic{equation}}
\setcounter{equation}{0}

We write all the non-metricity tensors for further use before
showing the comprehensive variation of $\mathcal{Q}$. They are given
as follows.
\begin{eqnarray}\label{B1}
\mathcal{Q}_{\phi\mu\nu}&=&\nabla_{\phi}g_{\mu\nu},
\\\label{B2}
\mathcal{Q}^{\phi}_{~\mu\nu}&=&g^{\phi\alpha}\mathcal{Q}_{\alpha\mu\nu}=
g^{\phi\alpha}\nabla_{\alpha}g_{\mu\nu}=\nabla^{\phi}g_{\mu\nu},
\\\label{B3}
\mathcal{Q}^{~~\mu}_{\phi~~\nu}&=&g^{\mu\beta}\mathcal{Q}_{\phi\beta\nu}=g^{\mu\beta}\nabla_{\phi}
g_{\beta\nu}=-g_{\beta\nu}\nabla_{\phi}g^{\mu\beta},
\\\label{B4}
\mathcal{Q}^{~~\nu}_{\phi\mu}&=&g^{\nu\beta}\mathcal{Q}_{\phi\mu\beta}=g^{\nu\beta}\nabla_{\phi}
g_{\mu\beta}=-g_{\mu\beta}\nabla_{\phi}g^{\nu\beta},
\\\label{B5}
\mathcal{Q}^{\phi\mu}_{~~~\nu}&=&g^{\phi\alpha}g^{\mu\beta}\nabla_{\alpha}g_{\beta\nu}
=g^{\mu\beta}\nabla^{\phi}g_{\beta\nu}=-g_{\beta\nu}\nabla^{\phi}g^{\mu\beta},
\\\label{B6}
\mathcal{Q}^{\phi~\nu}_{~\mu}&=&g^{\phi\alpha}g^{\nu\beta}\nabla_{\alpha}g_{\mu\beta}
=g^{\nu\beta}\nabla^{\phi}g_{\mu\beta}=-g_{\mu\beta}\nabla^{\phi}g^{\nu\beta},
\\\label{B7}
\mathcal{Q}^{~~\mu\nu}_{\phi}&=&g^{\mu\beta}g^{\nu\tau}\nabla_{\phi}g_{\beta\tau}
=-g^{\mu\beta}g_{\beta\tau}\nabla_{\phi}g^{\nu\tau}=-\nabla_{\phi}g^{\mu\nu},
\\\label{B8}
\mathcal{Q}^{\phi\mu\nu}&=&-\nabla^{\phi}g_{\mu\nu}.
\end{eqnarray}
Now, we calculate the variation of $Q$ by using Eq.(\ref{A8}) as
\begin{eqnarray}\nonumber
\delta\mathcal{Q}&=&-\frac{1}{4}\delta\big(-\mathcal{Q}^{\phi\nu\beta}\mathcal{Q}_{\phi\nu\beta}
+2\mathcal{Q}^{\phi\nu\beta}\mathcal{Q}_{\beta\phi\nu}-2\mathcal{Q}^{\beta}\tilde{\mathcal{Q}}_
{\beta} +2\mathcal{Q}^{\beta}\mathcal{Q}_{\beta}\big),
\\\nonumber
&=&-\frac{1}{4}\bigg(-\delta\mathcal{Q}^{\phi\nu\beta}\mathcal{Q}_{\phi\nu\beta}
-\mathcal{Q}^{\phi\nu\beta}\delta\mathcal{Q}_{\phi\nu\beta}
+2\delta\mathcal{Q}^{\phi\nu\beta}\mathcal{Q}_{\beta\phi\nu}
+2\mathcal{Q}^{\phi\nu\beta}\delta\mathcal{Q}_{\beta\phi\nu}
\\\nonumber
&-&2\delta\mathcal{Q}^{\beta}\tilde{\mathcal{Q}}_{\beta}
-2\mathcal{Q}^{\beta}\delta\tilde{\mathcal{Q}}_{\beta}
+\delta\mathcal{Q}^{\beta}\mathcal{Q}_{\beta}
+\mathcal{Q}^{\beta}\delta\mathcal{Q}_{\beta}\bigg),
\\\nonumber
&=&-\frac{1}{4}\bigg[\mathcal{Q}_{\phi\beta\tau}\nabla^{\phi}\delta
g^{\beta\tau}-\mathcal{Q}^{\phi\nu\beta}\nabla_{\phi}\delta
g_{\nu\beta}-2\mathcal{Q}_{\beta\phi\nu}\nabla^{\phi}\delta
g^{\nu\beta}+2\mathcal{Q}_{\phi\nu\beta}\nabla_{\beta}\delta
g_{\phi\nu}
\\\nonumber
&-&2\tilde{\mathcal{Q}}_{\beta}\delta(-g_{\mu\nu}\nabla^{\beta}g^{\mu\nu})
-2\mathcal{Q}^{\beta}\delta(\nabla^{\lambda}g_{\beta\lambda})
+\mathcal{Q}_{\beta}\delta(-g_{\mu\nu}\nabla^{\beta}g^{\mu\nu})
\\\nonumber
&+&\mathcal{Q}^{\beta}\delta(-g_{\mu\nu}\nabla_{\beta}g^{\mu\nu})\bigg],
\\\nonumber
&=&-\frac{1}{4}\bigg[\mathcal{Q}_{\phi\nu\beta}\nabla^{\phi}\delta
g^{\nu\beta}-\mathcal{Q}^{\phi\nu\beta}\nabla_{\phi}\delta
g_{\nu\beta}-2\mathcal{Q}_{\beta\phi\nu}\nabla^{\phi}\delta
g^{\nu\beta}+2\mathcal{Q}^{\phi\nu\beta}\nabla_{\beta}\delta
g_{\phi\nu}
\\\nonumber
&+&2\tilde{\mathcal{Q}}_{\beta}\nabla^{\beta}g^{\mu\nu}\delta
g_{\mu\nu}+2\tilde{\mathcal{Q}}_{\beta}g_{\mu\nu}\nabla^{\beta}\delta
g^{\mu\nu}-2\mathcal{Q}^{\beta}\nabla^{\lambda}\delta
g_{\beta\lambda}-\mathcal{Q}_{\beta}\nabla^{\beta}g^{\mu\nu}\delta
g_{\mu\nu}
\\\label{B9}
&-&\mathcal{Q}_{\beta}g_{\mu\nu}\nabla^{\beta}\delta
g^{\mu\nu}-\mathcal{Q}^{\beta}\nabla_{\beta}g^{\mu\nu}\delta
g_{\mu\nu}-\mathcal{Q}^{\beta}g_{\mu\nu}\nabla_{\beta}\delta
g^{\mu\nu}\bigg].
\end{eqnarray}
Using the following relations
\begin{eqnarray}\label{B10}
\delta g_{\mu\nu} &=&-g_{\mu\phi}\delta g^{\phi\alpha}g_{\alpha\nu},
\\\nonumber
&-&\mathcal{Q}^{\phi\nu\beta}\nabla_{\phi}\delta g_{\nu\beta},
\\\label{B11}
&=&-\mathcal{Q}^{\phi\nu\beta}\nabla_{\phi}\big(-g_{\nu\lambda}\delta
g^{\lambda\sigma}g_{\sigma\beta}\big),
\\\nonumber
&=&2\mathcal{Q}^{\phi\tau}_{~~\nu}\mathcal{Q}_{\phi\tau\mu}\delta
g^{\mu\nu}+\mathcal{Q}_{\phi\nu\beta}\nabla^{\phi}g^{\nu\beta},
\\\label{B12}
2\mathcal{Q}^{\phi\nu\beta}\nabla_{\beta}\delta g_{\phi\nu}
&=&-4\mathcal{Q}^{~~\tau\beta}_{\mu}\mathcal{Q}_{\beta\tau\nu}\delta
g^{\mu\nu}-2\mathcal{Q}_{\nu\beta\phi}\nabla^{\phi}\delta
g^{\nu\beta},
\\\label{B13}
-2\mathcal{Q}^{\beta}\nabla^{\lambda}\delta g_{\beta\lambda}
&=&2\mathcal{Q}^{\phi}\mathcal{Q}_{\nu\phi\mu}\delta
g^{\mu\nu}+2\mathcal{Q}_{\mu}\tilde{\mathcal{Q}}_{\nu}\delta
g^{\mu\nu}
\\\nonumber
&+&2\mathcal{Q}_{\nu}g_{\phi\beta}\nabla^{\phi}g^{\nu\beta},
\end{eqnarray}
Eq.(\ref{B9}) becomes
\begin{equation}\label{B14}
\delta \mathcal{Q}=2P_{\phi\nu\beta}\nabla^{\phi}\delta
g^{\nu\beta}-\big(P_{\mu\phi\alpha}\mathcal{Q}^{~~\phi\alpha}_{\nu}
-2\mathcal{Q}^{\phi\alpha}_{~~~\mu}P_{\phi\alpha\nu}\big)\delta
g^{\mu\nu},
\end{equation}
where
\begin{eqnarray}\nonumber
2P_{\phi\nu\beta}&=&-\frac{1}{4}\bigg[2\mathcal{Q}_{\phi\nu\beta}
-2\mathcal{Q}_{\beta\phi\nu}-2\mathcal{Q}_{\nu\beta\phi}
\\\label{B15}
&+&2(\tilde{\mathcal{Q}}_{\phi}-\mathcal{Q}_{\phi})g_{\nu\beta}+2
\mathcal{Q}_{\nu}g_{\phi\beta}\bigg],
\\\nonumber
4\big(P_{\mu\phi\alpha}\mathcal{Q}^{~~\phi\alpha}_{\nu}
-2\mathcal{Q}^{\phi\alpha}_{~~~\mu}P_{\phi\alpha\nu}\big)&=&
2\mathcal{Q}^{\phi\alpha}_{~~\nu}\mathcal{Q}_{\phi\alpha\mu}-4
\mathcal{Q}^{~~\phi\alpha}_{\mu}\mathcal{Q}_{\phi\mu\nu}
\\\nonumber
&+&2\tilde{\mathcal{Q}}^{\phi}\mathcal{Q}_{\nu\phi\mu}
+2\mathcal{Q}^{\phi}\mathcal{Q}_{\phi\mu\nu}
\\\label{B16}
&+&2\mathcal{Q}_{\mu}\tilde{\mathcal{Q}}_{\nu}-
\mathcal{Q}^{\phi}\mathcal{Q}_{\phi\mu\nu}.
\end{eqnarray}

\section*{Appendix C: Calculation of $\mathcal{Q}=-6H^{2}$}
\renewcommand{\theequation}{C\arabic{equation}}
\setcounter{equation}{0}

Using Eq.(\ref{A10}), we have
\begin{equation}\label{C1}
\mathcal{Q}=-\frac{1}{4}\bigg(-\mathcal{Q}_{\phi\mu\nu}
\mathcal{Q}^{\phi\mu\nu}+2\mathcal{Q}_{\phi\mu\nu}
\mathcal{Q}^{\mu\phi\nu}+\mathcal{Q}_{\phi}\mathcal{Q}^{\phi}
-2\mathcal{Q}_{\phi}\tilde{\mathcal{Q}^{\phi}}\bigg).
\end{equation}
Using the Appendix \textbf{B} for FRW metric, we obtain
\begin{eqnarray}\label{C2}
-\mathcal{Q}_{\phi\mu\nu}\mathcal{Q}^{\phi\mu\nu}&=&\nabla_{\phi}
g_{\mu\nu}\nabla^{\phi}g^{\mu\nu}=-12H^{2},
\\\label{C3}
\mathcal{Q}_{\phi\mu\nu}\mathcal{Q}^{\mu\phi\nu}&=&-\nabla_{\phi}
g_{\mu\nu}\nabla^{\mu}g^{\phi\nu}=0,
\\\label{C4}
\mathcal{Q}_{\phi}\mathcal{Q}^{\phi}&=&(g_{\beta\mu}\nabla_{\phi}
g^{\beta\mu})(g_{\tau\nu}\nabla^{\phi}g^{\tau\nu})=36H^{2},
\\\label{C5}
\mathcal{Q}_{\phi}\tilde{\mathcal{Q}^{\phi}}&=&(g_{\mu\beta}
\nabla_{\phi}g^{\mu\beta})(\nabla_{\alpha}g^{\phi\alpha}) =0.
\end{eqnarray}
Thus, we have
\begin{equation}\label{C6}
\mathcal{Q}=-\frac{1}{4}\bigg(-12H^{2}-0+36H^{2}+0\bigg)=-6H^{2}.
\end{equation}
\\
\textbf{Data Availability Statement:} No new data were created or
analyzed in this study.


\begin{thebibliography}{55}

\bibitem{1} Weyl, H.: Sitzungsber. Preuss. Akad. Wiss. \textbf{1}(1918)465.

\bibitem{3} Jimenez, J.B., Heisenberg, I. and  Koivisto, L.T.:
Phys. Rev. D \textbf{98}(2018)044048.

\bibitem{4} Linder, E. V.: Phys. Rev. D \textbf{82}(2010)109902.

\bibitem{5} Hammond, R.T.:  Rept. Prog. Phys. \textbf{65}(2002)599;
Arcos, H.I. and  Pereira, J.G.: Int. J. Mod. Phys. D
\textbf{13}(2004)2193.

\bibitem{6} Jimenez, J.B., Heisenberg, L. and Koivisto, T.:
Phys. Rev. D \textbf{98}(2018)044048.

\bibitem{6a} Sharif, M., Gul, M.Z.: Eur. Phys. J. Plus \textbf{133}(2018)345;
Chin. J. Phys. \textbf{57}(2019)329; Int. J. Mod. Phys. D
\textbf{28}(2019)1950054.

\bibitem{6b} Sharif, M. and Gul, M.Z.: Phys. Scr. \textbf{96}(2021)025002;
ibid. 125007; Adv. Astron. \textbf{2021}(2021)6663502; Eur. Phys. J.
Plus \textbf{136}(2021)503; Chin. J. Phys. \textbf{80}(2022)58; J.
Exp. Theor. Phys. \textbf{136}(2023)436; Gul, M.Z. and Sharif, M.:
Symmetry \textbf{15}(2023)684.

\bibitem{6c} Sharif, M. and Gul, M.Z.: Phys. Scr. \textbf{96}(2021)105001;
Pramana-J. Phys. \textbf{96}(2022)153; Universe \textbf{9}(2023)145.

\bibitem{6d} Sharif, M. and Gul, M.Z.: Int. J. Mod. Phys. A \textbf{36}(2021)2150004;
Universe \textbf{7}(2021)154; Chin. J. Phys. \textbf{71}(2021)365;
Mod. Phys. Lett. A \textbf{37}(2022)2250005; Int. J. Geom. Methods
Mod. Phys. \textbf{19}(2022)2250012.

\bibitem{6e} Sharif, M. and Gul, M.Z.: Fortschritte der Phys. \textbf{71}(2023)2200184;
Gen. Relativ. Gravit. \textbf{55}(2023)10; Phys. Scr.
\textbf{98}(2023)035030; Pramana-J. Phys. \textbf{97}(2023)122.

\bibitem{6f} Adeel, M. et al.: Mod. Phys. Lett. A \textbf{38}(2023)2350152;
Gul, M.Z. et al.: Eur. Phys. J. C \textbf{84}(2024)8; Rani, S. et
al.: Int. J. Geom. Methods Mod. Phys. \textbf{21}(2024)2450033.

\bibitem{7} Lu, J. et al.: Eur. Phys. J. C \textbf{79}(2019)530.

\bibitem{8} Jimenez, J.B., et al.: Phys. Rev. D \textbf{101}(2020)103507.

\bibitem{1a} Cai, Y.F. et al.: Class. Quantum Grav. \textbf{28}(2011)215011.

\bibitem{1b} Amani, A.R.: Int. J. Mod. Phys. D \textbf{25}(2016)1650071.

\bibitem{1c} Hohmann, M., Jarv, L. and Ualikhanova, U.: Phys. Rev. D \textbf{96}(2017)043508.

\bibitem{1d} Shabani, H. and Ziaie, A.H.: Eur. Phys. J. C \textbf{78}(2018)397.

\bibitem{1e} Sharif, M. and Saba, S.: Int. J. Mod. Phys. D \textbf{28}(2019)1950077;
J. Exp. Theor. Phys. \textbf{128}(2019)571.

\bibitem{1f} Bhattacharjee, S. and Sahoo, P.K.: Phys. Dark Universe \textbf{28}(2020)100537.

\bibitem{1g} Bhardwaj, V.K. et al.: Can. J. Phys. \textbf{100}(2022)475.

\bibitem{2a} Lazkoz, R. et al.: Phys. Rev. D \textbf{100}(2019)13219.

\bibitem{2d} Mandal, S. et al.: Phys. Rev. D \textbf{102}(2020)024057.

\bibitem{2c} Mandal, S., Wang, D. and Sahoo, P.K.: Phys. Rev. D \textbf{102}(2020)124029.

\bibitem{2e} Bajardi, F., Vernieri, D. and Capozziello, S.:
Eur. Phys. J. Plus \textbf{135}(2020)912.

\bibitem{2f} Mandal, S. et al.: Eur. Phys. J. Plus \textbf{136}(2021)760.

\bibitem{27} Harko, T. et al.: Phys. Rev. D \textbf{98}(2018)084043.

\bibitem{28} Cai Y.F. and Easson D.A.: Astropart. Phys. \textbf{8}(2012)020.

\bibitem{29} Bamba, K. et al.: J. Cosmol. Astropart. Phys. \textbf{01}(2014)008;
Nojiri, S., Odintsov, S.D. and Oikonomou, V.K.: Phys. Rev. D
\textbf{93}(2016)084050.

\bibitem{26} Singh, J.K. et al.: J. High Energy Phys. \textbf{2023}(2023)21.

\bibitem{30} Raychaudhuri, A.: Phys. Rev. \textbf{98}(1955)1123; Ehlers, J.:
Int. J. Mod. Phys. D \textbf{15}(2006)1573.

\bibitem{31} Hawking, S.W. and Ellis, G.F.R.: \emph{The Large Scale Structure
of Spacetime} (Cambridge University Press, 1973).

\bibitem{32} Ilyas, M. and Rehman, U.W.: Eur. Phys. J. C \textbf{81}(2020)160.

\bibitem{33} Sandage, A.: Astrophys. J. \textbf{136}(1962)319.

\bibitem{34} Capozziello, S. et al.: Phys. Lett. B \textbf{832}(2022)137229.

\bibitem{35} Starobinsky, A.A.: J. Exp. Theor. Phys. Lett. \textbf{68}(1998)757;
Huterer, D. and Turner, M.S.: Phys. Rev. D \textbf{60}(1999)081301;
Lazkoz, R., Nesseris, S. and Perivolaropoulos, L.: J. Cosmol.
Astropart. Phys. \textbf{11}(2005)010.

\bibitem{36} Rubakov, V.A.: Phys. Usp. \textbf{57}(2014)128; Creminelli, P. et al.:
J. High Energy Phys. \textbf{2006}(2006)080.

\bibitem{37} Aghanim, N. et al.: Astron. Astrophys. \textbf{641}(2020)A6.

\end{thebibliography}
\end{document}